# Speed Control of Multi Level Inverter Designed DC Series Motor with Neuro-Fuzzy Controllers

G.MadhusudhanaRao[1], Dr. B.V.SankerRam[2]
[1] Dept. of EEE, JNTU-Hyderabad, India
[2] Dept. of EEE, JNTU-Hyderabad, India

***Abstract:*** This paper describes the speed control of a DC series motor for an accurate and high-speed performance. A neural network based controlling operation with fuzzy modeling is suggested in this paper. The driver units of these machines are designed with a Multi-level inverter operation and are controlled by a common current control mechanism for an accurate and efficient driving technique for DC series motor. The neuro-fuzzy logic control technique is introduced to eliminate uncertainties in the plant parameters of the DC Series motors, and also considered as potential candidate for different applications to prove adequacy of the proposed control algorithm through simulations. The simulation result with such an approach is made and observed efficient over other controlling technique.

## I. INTRODUCTION

With the increased demand for higher load operative system, DC motors are coming out to be effective solutions, due to their high torque supporting nature and robust controlling operation. With the need for DC machine operation the need for robust speed controller is in great requirement. Various speed controllers have been suggested in the past for the improvement of speed controlling in DC motor operation. The DC motor is the obvious proving ground for advanced control algorithms in electric drives due to the stable and straightforward characteristics associated with it. It is also ideally suited for trajectory control applications as shown in reference [1-3]. From the control systems point of view, the DC motor can be considered as SISO plant, thereby eliminating the complications associated with a multi-input drive system. The system can be represented as shown in figure.1which is known as mathematical model of dc motor. There are various approaches made in past to control the speed of this machine model either by providing internal or external controlling mechanisms. These approaches are found to be more accurate in the approach of controlling the speed by the current control strategies. A common approach of applying a feedback controlling strategy is shown in figure 2.

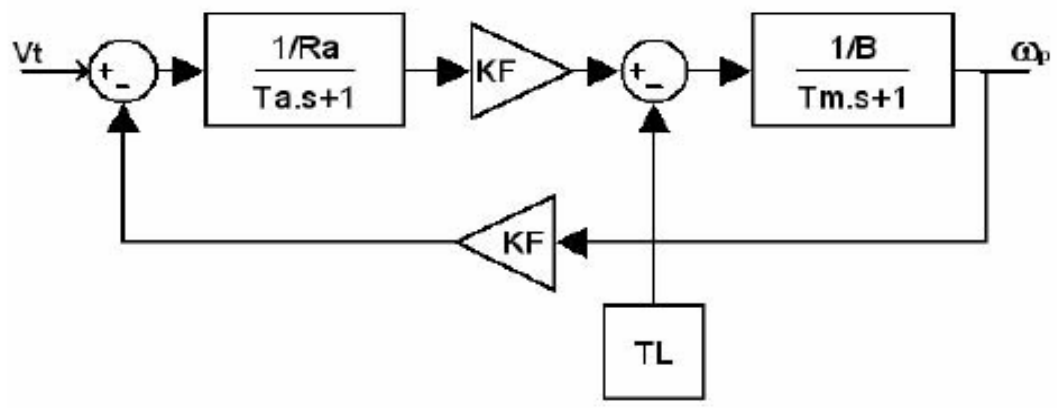

Figure 1. The mathematical model of a DC motor

There are different methods been made to develop a control system for DC motor, a conventional control system of DC motor, where the regulator current and regulator speed are synthesized by Bietrage-optimum to reduce the over-regulation [6] is used as the analyzing system.

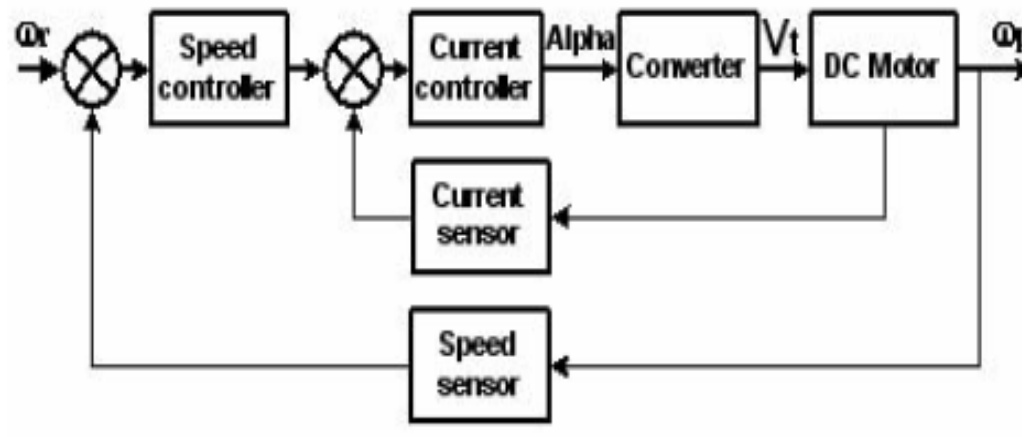

Figure 2. A conventional current control strategy for DC motor

Where various approaches were made in past to control the speed of a DC machines based on the above show architecture all the conventional methods uses a 3-current measurements for the controlling of the DC machine speed. These techniques hence require higher and complex controlling strategy and are low robust to the variations. These techniques also demands for high maintenance cost for the current measuring sensors. To overcome the issue in current DC motor speed controller in this paper a novel approach of controlling the speed of a DC machine by a common current scheme is suggested. To make an appropriate decision of the controlling signal to the inverter neuro-fuzzy decision architecture is incorporated to the feedback circuitry for providing accurate controlling signal to the inverter unit. Further to provide a stability under variable input condition in this paper an approach to designing an multi level inverter in DC driving circuitry is suggested.

## II. CONTROL STRATEGY

In this paper a simple and efficient modulation control system, which allows to have good current waveform. To fulfill the objective of accurate speed controlling is suggested for a DCM. The DCM has the advantage of;

1. The position sensor system for the shaft needs only to deliver six digital signals for commanding the transistor of the inverter.
2. The quasi square-wave armatures current are mainly characterized by their maximum amplitude value, which directly controls the machine torque.
3. The inverter performance is very much reliable because there are natural dead times for each transistor.

The first and third characteristic allows reducing the complex circuitry required by other machines and allows the self synchronization process for the operation of the machine. The second characteristic allows designing a circuit for controlling only a dc component which represents the maximum amplitude value of the trapezoidal current, $I_{MAX}$

The most popular way to control DCM for location application is through voltage-source current controlled inverters. The phase currents, torque and speed can be adjusted $T = K_T\ I_{MAX,}$ $K_T$ = torque constant. There are two ways to control the phase current of a DCM.

1. Through the measurement of the phase currents which are compared and forced to follow a quasi square template.
2. Through the measurement of the dc link current, $I_{MAX}$

In the first case, the control is complicated because it is required to generate three quasi square current templates shifted 120º for the three phases. In the second case, it is difficult to measure the dc current, because the connection between transistor and the dc capacitors in power inverters is made with flat plates to reduce leakage inductance. Then, it becomes difficult to connect a current sensor. To avoid those drawbacks, a equivalent dc current can be obtained through the sensing of the armature currents. These currents are rectified and a dc component, which corresponds to the amplitude $I_{MAX}$ of the original phase current, is obtained. These obtained current values can be passed to a neuro-fuzzy controller for the speed controlling operation.

## III. NEURO-FUZZY CONTROLLING APPROACH

Neural network have been found to be effective systems for learning discriminates for patterns from a body of examples [5]. Activation signals of nodes in one layer are transmitted to the next layer through links which either attenuate or amplify the signal. ANNs are trained to emulate a function by presenting it with a representative set of input/output functional patterns. The back-propagation training technique adjusts the weights in all connecting links and thresholds in the nodes so that the difference between the actual output and target output are minimized for all given training patterns [1]. It is also widely accepted that maximum of two hidden layers are sufficient to learn any arbitrary nonlinearity [4]. However, the number of hidden neurons and the values of learning parameters, which are equally critical for satisfactory learning, are not supported by such well established selection Criteria.

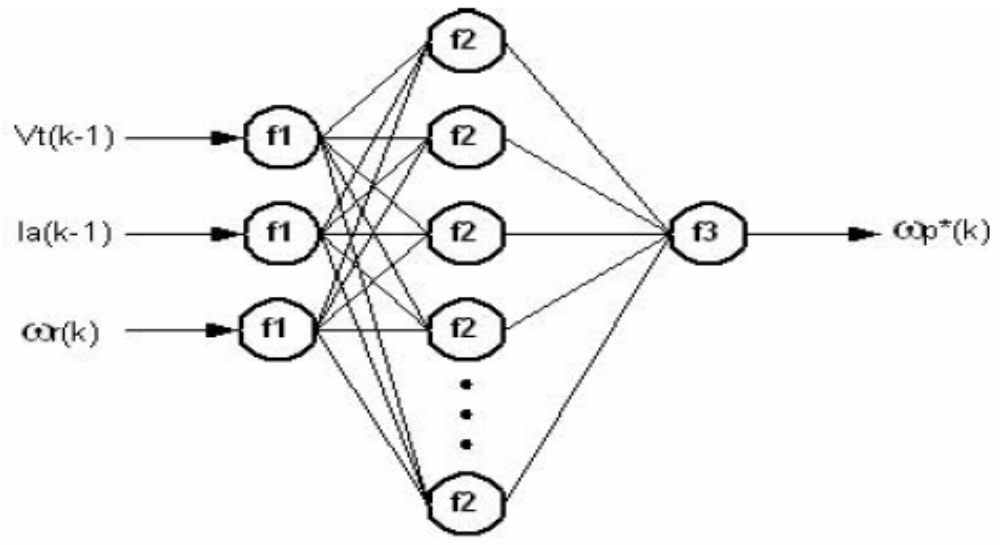

f1: tansig; f2:tansig; f3: purelin

Figure 3. Structure of ANN1

The ANN1 and ANN2 structure is shown in Figure 3, and Figure 4. It consists of an input layer, output layer and one hidden layer. The input and hidden layers are tansig-sigmoid activation functions, while the output layer is a linear function. Three inputs of ANN1 are a reference speed ωr(k), a terminal voltage Vt(k-1) and an armature current ia(k-1). And output of ANN1 is an estimated speed ωp*(k). The ANN2 has four inputs: reference speed ωr(k), a terminal voltage Vt(k-1), an armature current ia(k-1) and an estimated speed ωp*(k) from ANN1. The output of ANN2 is the control signal for converter Alpha.

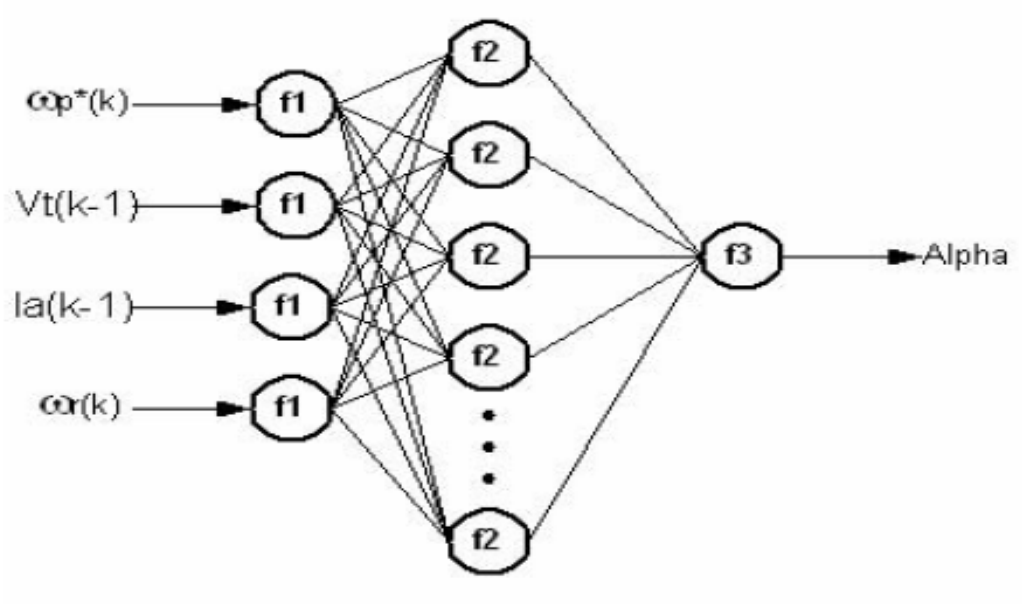

f1: tansig; f2:tansig; f3: purelin
Figure 4. Structure of ANN2

The obtained decision and the designed neural network are observed to be higher in decision accuracy but takes larger computation time for the error convergence to give an estimation decision. To improve the decision speed by the network fuzzy decision architecture in incorporated with the neural network. The fuzzy modeling is done by using the characteristic function defined as

$$\mu_Y(x) = \begin{cases} 1 \text{ if } x \in Y \\ 0, \text{ otherwise} \end{cases}$$

The subset Y can be uniquely represented by ordered pairs Y = {($x_1$, 1), ($x_2$, 1), ($x_3$, 0), ($x_4$, 0), ($x_5$, 1)}. The second member of an ordered pair called as the membership grade of the appropriate element can take its value not only from the set {0, 1} but from the closed interval [0, 1] as well. For a X a universal crisp set (observation set). The set of ordered pairs Y = {(x, $\mu_Y$ (x))|x ∈X, $\mu_Y$ : X [0, 1]} is said to be the fuzzy subset of X. The $\mu_Y$: X [0, 1] function is called as membership function and its value is said to be the membership grade of x.

A if-then rule is used for the decision approach for the input observation in fuzzy model. **if-then** rules, $c_{ik}$(k = 0, 1, . . . , n) are the consequent parameter resulting in $y_i$ output from the $i^{th}$ if-then rule, for $X_k$ is a fuzzy set. Given an input ($x_1$, $x_2$, . . ., $x_n$), the final output of the fuzzy model is referred as follows:

$$Y = \frac{\sum_{i=1}^{N} \omega i y i}{\sum_{i=1}^{N} \omega i}$$

$$= \frac{\sum_{i=1}^{N} \omega (Ci1 + Ci1T1 + \cdots \ldots Ci1Tn)}{\sum_{i=1}^{N} \omega i}$$

$$= \frac{\sum_{i=1}^{N} \; \sum_{i=1}^{N} \omega CikTk}{\sum_{i=1}^{N} \omega i}$$

Where $x_0$ = 1, $\omega_i$ is the weight of the $i^{th}$ If-Then rule for the input and is calculated as,

$$\omega i = \prod_{k1}^{n} Aik(xk)$$

$A_{ik}(x_k)$, where $A_{ik}(x_k)$ is the grad of membership of $X_k$. If $x_1$ is $A_{i1}$ . . . , $x_n$ is $A_{in}$ then $y_i = k_i(c_0 + c_1x_1 + \ldots + c_nx_n)$ where i = 1, 2, . . .,N, N is the number of if-then rules. The advantage of solving the complex nonlinear problems by utilizing fuzzy logic methodologies is that the experience or expert's knowledge described as a fuzzy rule base can be directly embedded into the systems for dealing with the problems. This advantage of the fuzzy rules help in making the network designed to take more accurate decision based on the selected information by fuzzy node resulting in faster convergence of the neural network. Basic model of such an neuro fuzzy architecture is as shown below in figure 5.

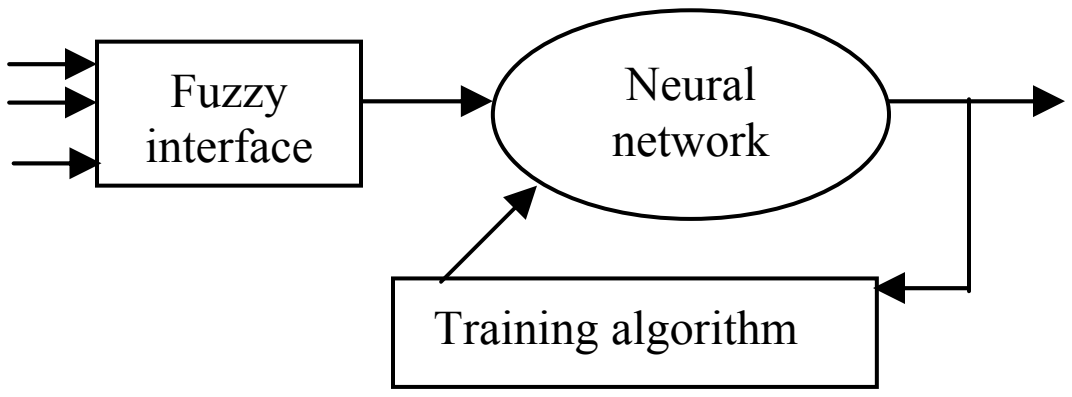

Figure 5. A Neuro-fuzzy controlling unit

The obtained control signals from this neuro fuzzy logic model is then passed to a inverter model design with a multi level inverter logic.

# IV. MULTI-LEVEL INVERTER DESIGN

For the driving unit of the DCM a Sinusoidal PWM (SPWM) is developed. SPWM for Multilevel Inverter is based on classic two levels SPWM with triangular carrier and sinusoidal reference waveform as shown in figure 6.

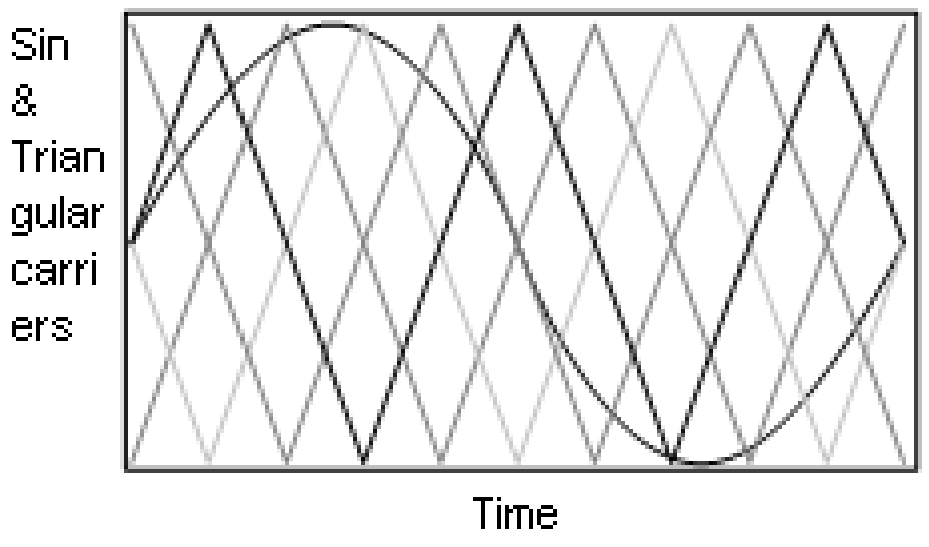

(a)

Figure 6 (a) vertically shifted carriers

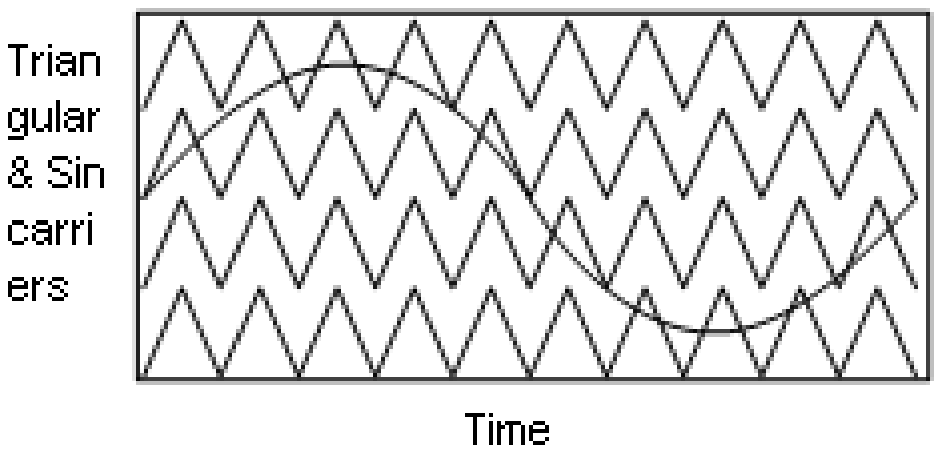

(b)

Figure 6 (b) vertically shifted carriers

Only difference between two level SPWM and multilevel SPWM is, numbers of carriers are used in multilevel SPWM. For '*m*' level inverter '*m*-1' carrier are used. Carriers used in multilevel inverter may be vertically shifted or horizontally shifted as shown in Fig 6(a),(b). Vertically shifted carrier scheme can be more easily implemented on any digital controller. A vertically shifted scheme comes with three variants,

1) All carriers are in phase (PH disposition)

2) All carries above the zero reference are in phase, but in opposition with those below (PO disposition)

3) All carriers are alternatively in opposition (APO disposition)

The PH technique produce less harmonics on a line-to-line basis compared to other two techniques For five level inverter, four carriers ($C_1 - C_4$) divides whole modulating voltage in to four region $r_1$ to $r_4$ as shown in Fig 6 (a).

Implementation of the SPWM technique is based on classical SPWM technique with carriers and reference sine waveform. Reference wave form in digital SPWM represents a sample and hold waveforms of sine wave form.

This sampling of sine waveform comes in two variants; a) Symmetrical sampling, b) Asymmetrical sampling.

In symmetrical sampling, reference sine waveform is sampled at only positive peak of the carrier waveform and sample is held constant for the complete carrier period. Here sampling frequency is equal to carrier frequency. The phase shift is given by $\frac{\pi}{m_f}$, where

$$m_f = \frac{f_c}{f_m}$$

$f_c$ = Carrier frequency .

$f_m$ = Reference Sine wave frequency.

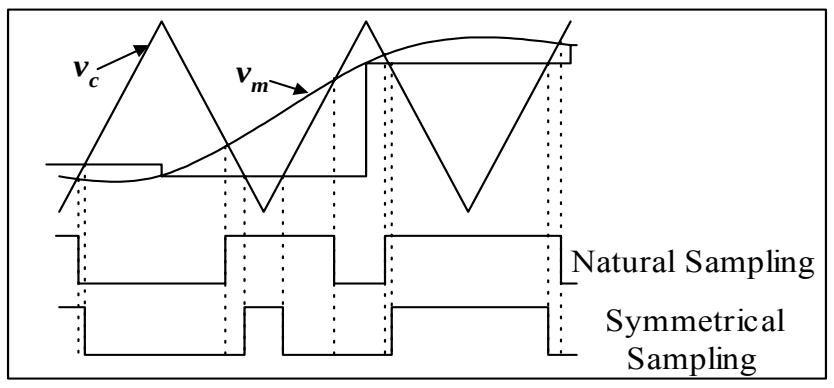

Figure 7. Natural sampling, Symmetrical Sampling

In asymmetrical sampling, the reference signal is sampled at positive as well as negative peak of carrier frequency and held constant for half the carrier period. Here sampling frequency is twice the carrier frequency. Asymmetrical sampling is the preferred method, since each switching edge is the result of new sample and gives better performance as shown in Fig 4. The phase shift is by $\frac{\pi}{2m_f}$.

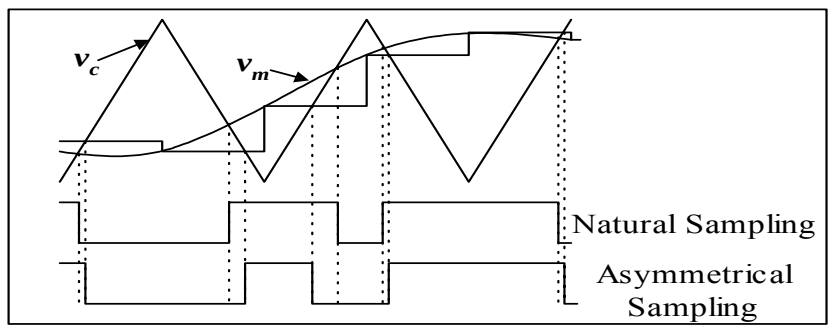

Figure 8. Natural sampling, asymmetrical sampling.

Comparing natural SPWM and digital SPWM, this digital SPWM has the following disadvantages,

1) Digital SPWM method sample the signal input at the beginning of the switch cycle, before the actual switching edge reflects this value later in the cycle.

2) This introduces a delay in out-put waveform. A delay of $\frac{\pi}{m_f}$ and $\frac{\pi}{2m_f}$ is introduced in symmetrical and asymmetrical sampling method respectively, where $m_f$ is frequency modulation ratio

3) This delay in response is significant when the ratio of switch frequency to reference frequency (the pulse number) is small. It leads to a frequency response roll-off which obeys a Bessel function, similar to the familiar sine function roll-off for Pulse Amplitude Modulation (PAM).
4) Another unwanted effect of digital SPWM is odd harmonic distortion of the synthesized waveform. The severity of these effects is a function of the ratio of the modulating and carrier frequencies, $f_1/f_c$. This ratio may approach and pass unity in high power active filters (high $f_1$, low $f_c$), by which point these effects have become significant and limiting. In proposed model, magnitude of modulating signal at crossover instant is calculated at interval of $T_s/2$ at each peak of carrier frequency. $k^{th}$ sample give the value of the discrete time signal $t_k = kT_s/2$ where $k$ is integer. Extrapolation process is carried out to find the intersection of modulating signal.

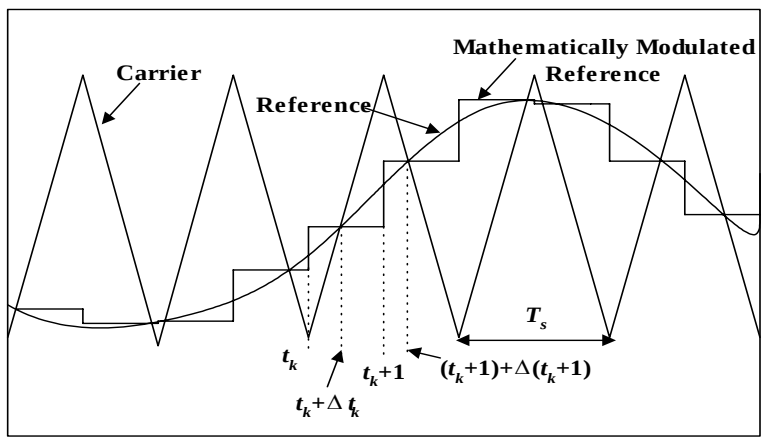

Figure 9. Scheme for proposed SPWM method.

As shown in Figure 9 there is time delay $\Delta t_k$ between sampling instant $t_k$ and actual crossing of natural sine waveform and triangular carrier waveform $t_k+\Delta t_k$. Because of this time delay their lies a phase delay in output waveform as shown in Figure 8. If this time delay $\Delta t_k$ can be calculated then instead of using sampled value of sine waveform at time instant $t_k$ for comparing with carrier, a sampled value of sine waveform at $t_k+\Delta t_k$ can be used. Procedure of calculating this time delay $\Delta t_k$ is as follows, Consider reference signal as,

$$V_r(t) = m_a V_m \sin(\omega_m t)$$

Where
$m_a$ = modulation index .
$V_m$ = Peak value of Reference signal .
$\omega_m = 2\,\pi f_m$ .
$f_m$ = fundamental frequency of reference signal .
$t_k$ = Time instant at which sine wave form is sampled.
Carrier signal equation for positive slop and negative slop,

$$V_{c(P_S)} = 2V_c f_c t - \frac{V_c}{2}$$

$$V_{c(N_S)} = -2V_c f_c t + \frac{V_c}{2}$$

$V_c$ = Peak value of carrier signal.
$f_c$ = Frequency of carrier signal .
The value of '$\Delta t_k$' can be found simply by equating values of reference signal and rising edge (positive slop) of carrier signal at instant of intersection (i.e. $t_k+\Delta t_k$), and of reference signal and falling edge (negative slop) of carrier signal at instant of intersection (i.e. $t_{(k+1)}+\Delta t_{(k+1)}$). The allocation of proposed mathematical model can be extended to multilevel inverter [5] [6] [7]. The only difference in above procedure and procedure for determination $\Delta t_k$ in case of multilevel inverter is that, as numbers of carriers are used in multilevel inverter, exact region of interaction of reference and carrier is to be known[8][9]. Figure 10 shows the reference and carrier waveform arrangements necessary to achieve PD SPWM for a five level inverter. Each shifted carrier is considering as one region.

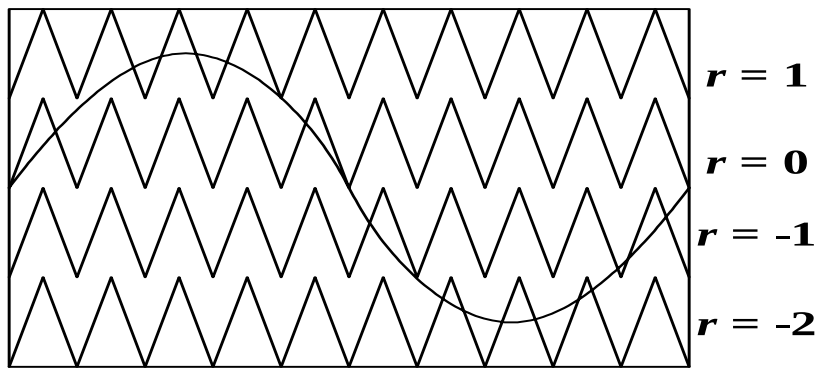

Figure 10. Distribution of regions for proposed SPWM

Transition from one region of operation to the other can be decided on the basis of calculated vale of '$\Delta t_k$'. To decide the transition from one region to other the criterion of transition for positive slop carrier cross-over is,
(1) If $\Delta tk > 1/2f_c$, then transition is form lower region to upper region, so $r_{new} = r_{old} + 1$ (where r = region).
(2) If $\Delta t_k < 0$, then transition is from upper region to lower region, so $r_{new} = r_{old} - 1$.
Similarly, to decide the transition from one region to other the criterion for negative slop carrier cross-over is,
1) If $\Delta t_k > 1/2f_c$, Then transition is form upper region to lower region, so $r_{new} = r_{old} - 1$.
2) If $\Delta t_k < 0$ Then transition is from lower region to upper region, so $r_{new} = r_{old} + 1$.

## V. SIMULATION OBSERVATIONS

The proposed approach is developed over a generalized controlling architecture of a DCM as illustrated in figure 11.

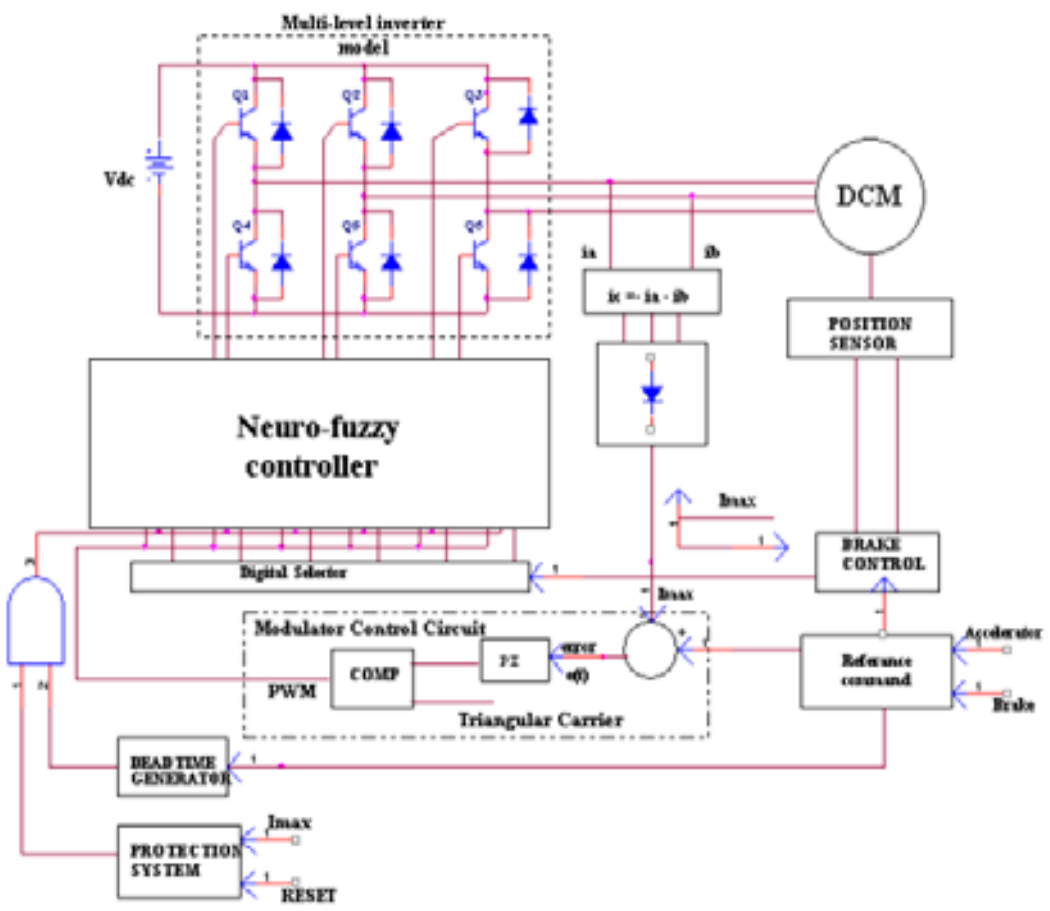

Figure 11. Simulation model for proposed DCM controller architecture

Different modulation scheme for the suggested controller units are carried out at a) Selective current distortion and at b) SPWM method. In SPWM method of modulation for multilevel inverter numbers of carriers are used. Arrangements of these carriers come with different variants. Fig. 8 gives (a) carrier arrangement, (b) output voltage and (c) FFT for PH disposition (All carriers are in phase) SPWM method for 5-level inverter. ($f_c$ = 1050 Hz, $f_m$ = 50 Hz).

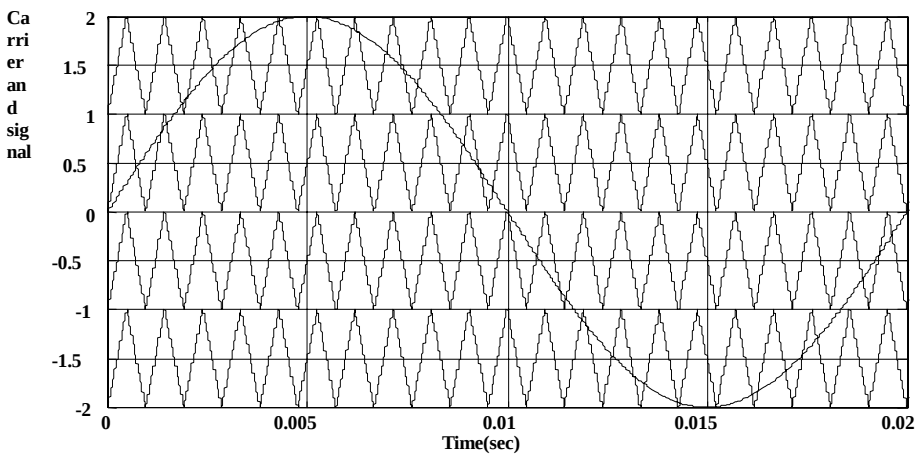

(a)

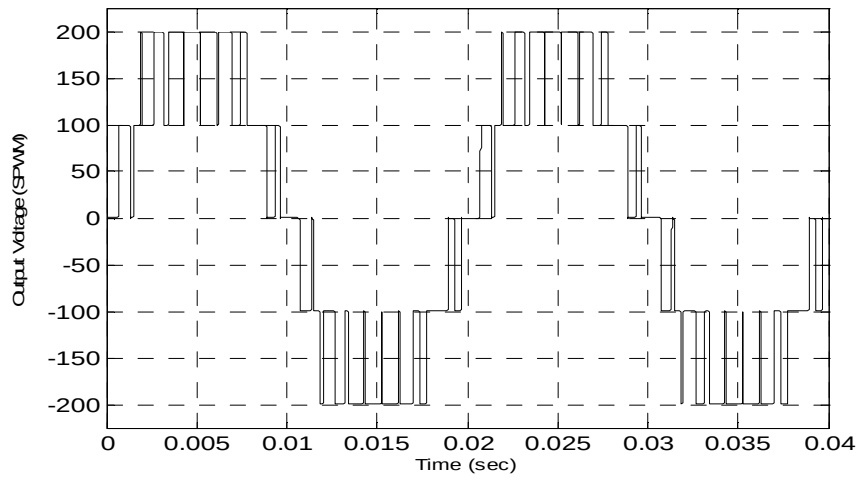

(b)

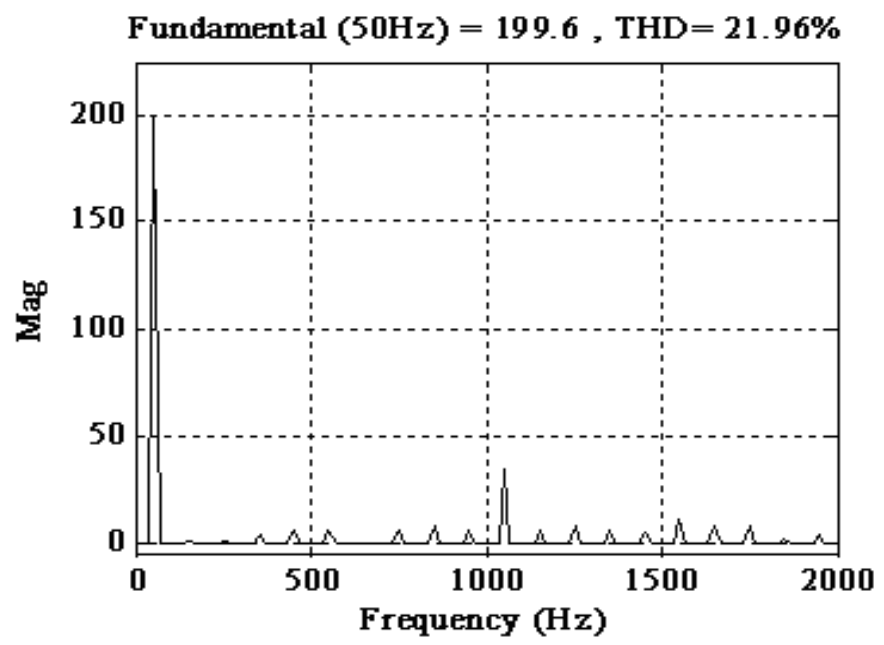

(c)

Figure.12 (a) carrier arrangement, (b) output voltage and (c) FFT for PO disposition (All carries above the zero reference are in phase, but in opposition with those below ) SPWM method for 5-level inverter model.($f_c$ = 1050 Hz, $f_m$ = 50 Hz).

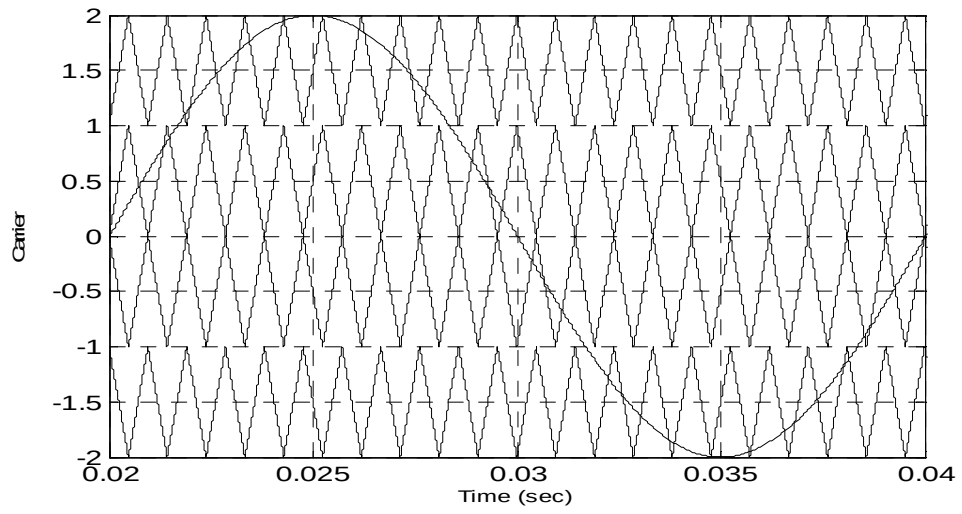

(a)

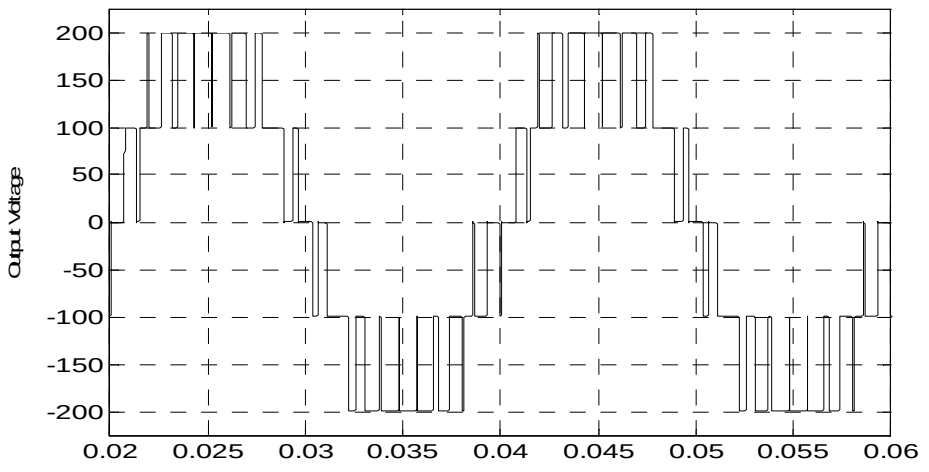

(b)

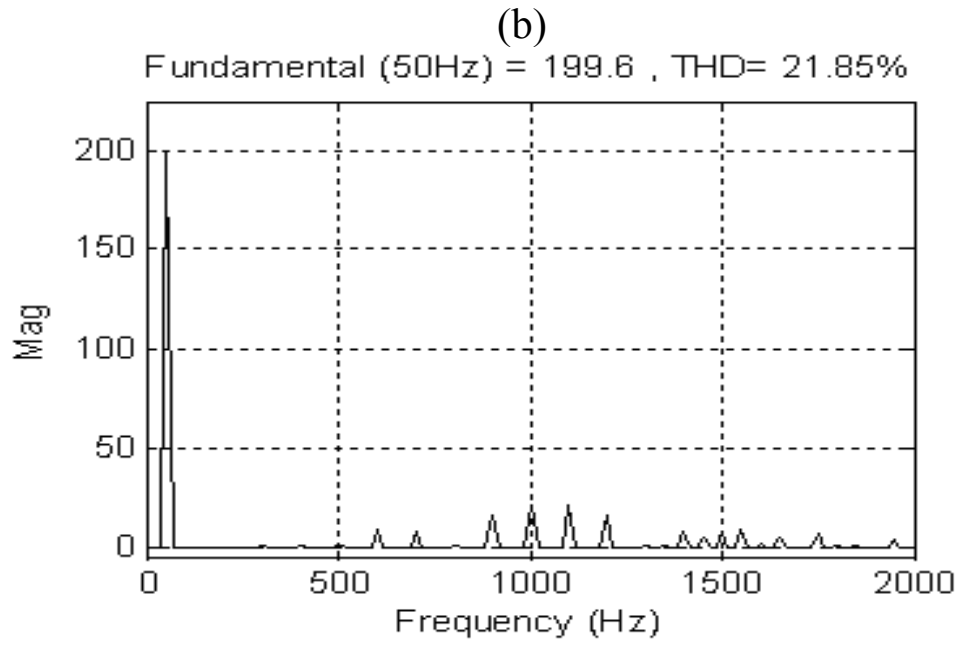

(c)

Figure. 13. (a) carrier arrangement, (b) output voltage and (c) FFT for APO disposition (All carriers are alternatively in opposition) SPWM method for 5-level inverter model.($f_c$ = 1050 Hz, $f_m$ = 50 Hz)

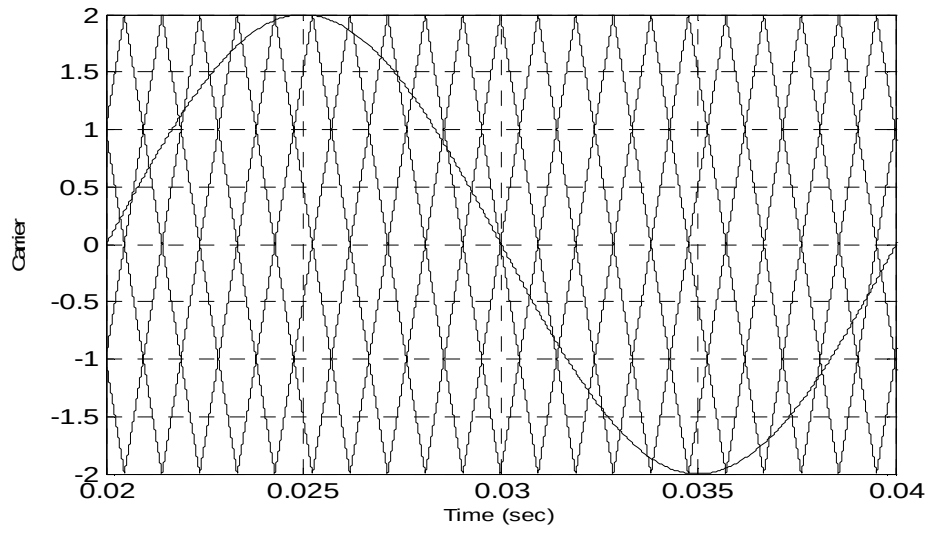

(a)

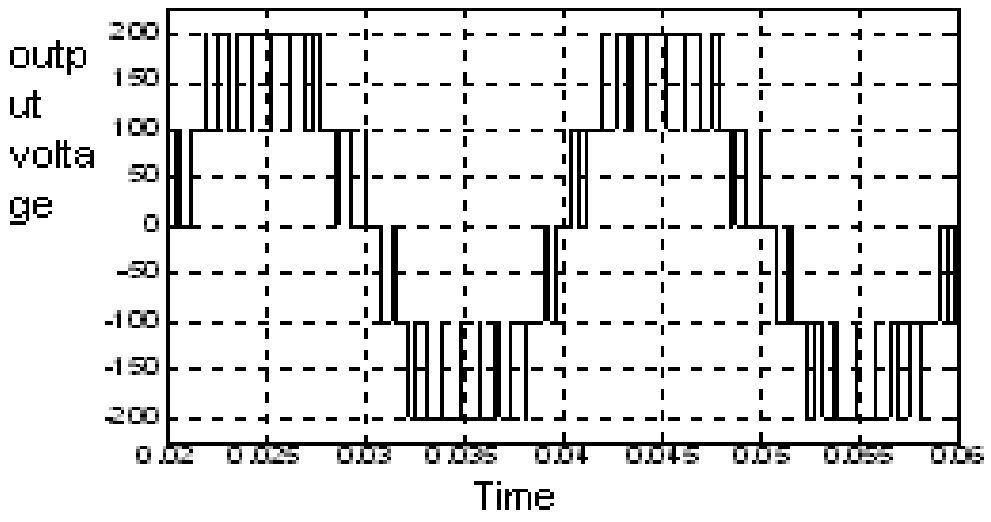

(b)

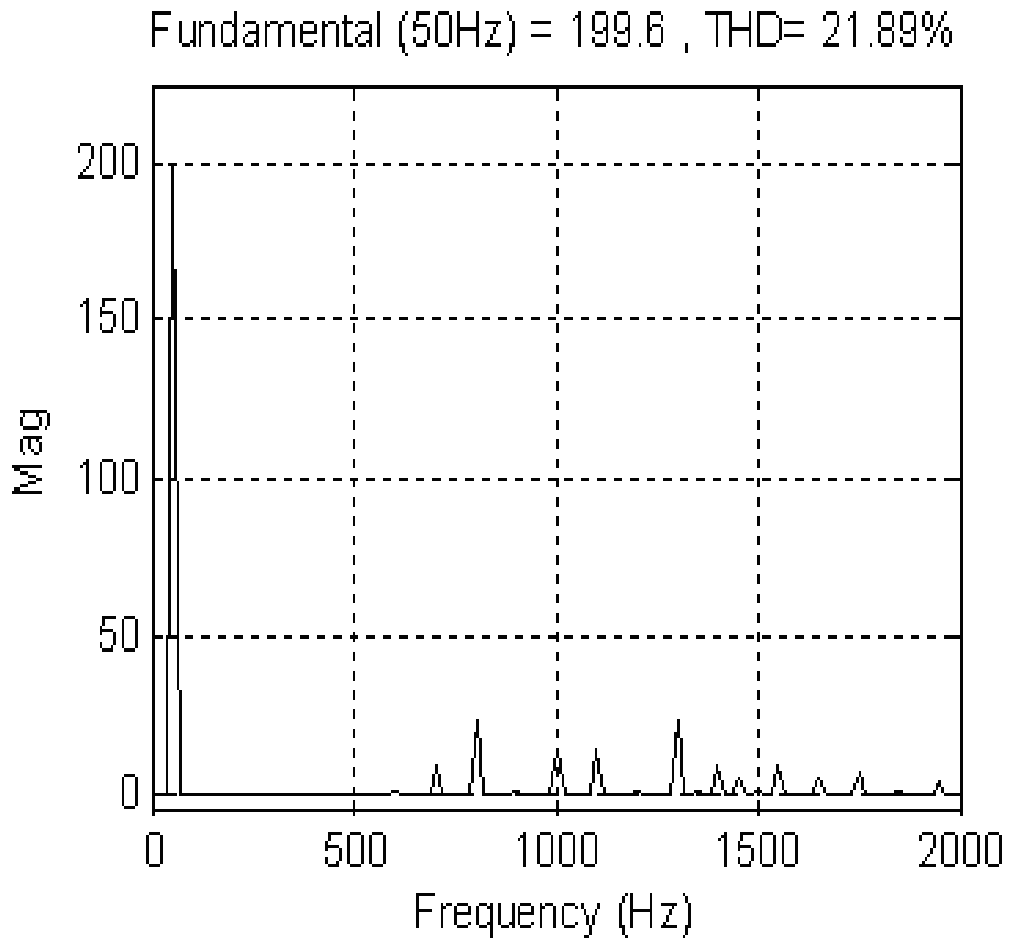

(c)

Figure. 14 (a) carrier arrangement, (b) output voltage and (c) FFT for SPWM method for 5-level inverter where carriers are shifted by 90° with respective to each other. ($f_c$ = 1050 Hz, $f_m$ = 50 Hz)

TABLE.1
COMPARISON DIFFERENT SPWM METHODS FOR MULTILEVEL INVERTER.

| Comparison of SPWM method for Multilevel inverter | | |
|---|---|---|
| | Method | THD (%) |
| 1 | PH | 21.96 |
| 2 | PO | 21.85 |
| 3 | APO | 21.89 |
| 4 | Carrier shift (90°) | 21.28 |

# VI. CONCLUSION

An approach for the speed controlling of DC series machine based on the neural network and fuzzy architecture is developed. The system is incorporated with a multilevel inverter design for the accurate controlling of the driving current to the DC series motor. A simulation model for the stated approached is developed with Mat lab/Simulink design and the performance for developed control strategy is presented. It is observed that the current distortion is minimized to the input circuitry by the incorporation of a multilevel inverter. A neuro-fuzzy controller into the driving circuitry provides better performance than the conventional one. The neruo-fuzzy model with multi level inverter circuitry is observed to be providing higher speed stability and disturbance reduction as compared to the conventional feedback control logics.

A different control mechanism for dc machines has been simulated and presented using measurement approaches. It is based on the generation of quasi-square currents using only one current controller for the three phases.

These characteristics allow using the triangular carrier as a current control mechanism for the power transistors, which is simpler and more accurate than other options. This control mechanism with ANN has been compared with conventional techniques which showed the excellent characteristic with the above modified approach of the DC series motor has been successfully controlled using an ANN.